\documentclass[review]{elsarticle}

\journal{Journal of \LaTeX\ Templates}

\begin{document}

\def\rd{{\rm d}}
\newcommand{\blue}[1]{\textcolor{blue}{#1}}
\newcommand{\red}[1]{\textcolor{red}{#1}}
\newcommand{\green}[1]{\textcolor{green}{#1}}
\def\vf {{\bf f}}
\def\vj {{\bf j}}
\def\vn {{\bf n}}
\def\vP{{\bf p}}
\def\vp{{\bf p}}
\def\vx{{\bf x}}
\def\vu{{\bf u}}
\def\vv{{\bf v}}
\def\vw {{\bf w}}
\def\mA{{\bf A}}
\def\mB{{\bf B}}
\def\mC{{\bf C}}
\def\mD{{\bf D}}
\def\mG{{\bf G}}
\def\mI{{\bf I}}
\def\mS{{\bf S}}
\def\mQ{{\bf Q}}
\def\mR{{\bf R}}
\def\mT{{\bf T}}
\def\mU{{\bf U}}
\def\mV{{\bf V}}
\def\w{\omega}
\def\u{\nu}
\def\mGa{\mbox{\boldmath$\Gamma$}}
\def\mPhi{\mbox{\boldmath$\Phi$}}
\def\mPi{\mbox{\boldmath$\Pi$}}
\def\mXi{\mbox{\boldmath$\Xi$}}
\def\vg {\mbox{\boldmath$\gamma$}}
\def\vpi{\mbox{\boldmath$\pi$}}
\def\vphi{\mbox{\boldmath$\phi$}}
\def\vnu{\mbox{\boldmath$\nu$}}
\def\vkappa{\mbox{\boldmath$\kappa$}}
\def\vmu{\mbox{\boldmath$\mu$}}
\def\wta {\widetilde{a}}
\def\wtb {\widetilde{b}}
\def\wtx {\widetilde{x}}

\begin{frontmatter}

\title{A framework towards understanding mesoscopic phenomena:
Emergent unpredictability, symmetry breaking and dynamics across scales}

\author[add1]{Hong Qian}

\author[add2]{Ping Ao}

\author[add3]{Yuhai Tu}

\author[add4]{Jin Wang\corref{corr}}
\cortext[corr]{Corresponding author}
\ead{jin.wang.1@stonybrook.edu}

\address[add1]{Department of Applied Mathematics,University of Washington, Seattle, WA 98195 USA}
\address[add2]{Shanghai Center for Systems Biomedicine,Shanghai Jiao Tong University, 200240, Shanghai, PRC}
\address[add3]{IBM T.J. Watson Research Center, Yorktown Heights, NY 10598 USA}
\address[add4]{Department of Chemistry and Physics,State University of New York,Stony Brook, NY 11794 USA.}

\begin{abstract}
By integrating four lines of thoughts: symmetry breaking
originally advanced by Anderson, bifurcation from nonlinear
dynamical systems, Landau's phenomenological theory of
phase transition, and the mechanism of emergent rare
events first studied by Kramers, we introduce a
possible framework for understanding {\em mesoscopic dynamics} that links ($i$) fast microscopic (lower level) motions,
($ii$) movements within each basin-of-attraction at
the mid-level, and ($iii$) higher-level rare transitions
between neighboring basins, which have slow
rates that decrease exponentially with the size of the
system.  In this mesoscopic framework, the fast dynamics is
represented by a rapidly varying stochastic process
and the mid-level by a nonlinear dynamics.
Multiple attractors arise as emergent properties of the
nonlinear systems.
The interplay between the stochastic element and
nonlinearity, the essence of Kramers' theory, leads
to successive jump-like transitions among different
basins.  We argue each transition is a dynamic
symmetry breaking, with the potential of
exhibiting Thom-Zeeman catastrophe
as well as phase transition with the breakdown of ergodicity (e.g., cell differentiation).  The slow-time dynamics of the
nonlinear mesoscopic system is not deterministic,
rather it is a discrete stochastic jump process.
The existence of these discrete states and the Markov transitions among them
are both emergent phenomena.  This emergent
stochastic jump dynamics then serves as the stochastic
element for the nonlinear dynamics of a higher level
aggregates on an even larger spatial and slower time
scales (e.g., evolution).  This description
captures the hierarchical structure outlined by Anderson
and illustrates two distinct types of limit of a mesoscopic
dynamics: A long-time ensemble thermodynamics in terms
of time $t\rightarrow\infty$ followed by the size of the
system $N\rightarrow\infty$, and a short-time trajectory
steady state with $N\rightarrow\infty$ followed by $t\rightarrow\infty$. With these limits, symmetry
breaking and cusp catastrophe are two perspectives of
the same mesoscopic system on different time scales.
\end{abstract}


\end{frontmatter}



\section{Introduction}

	There is a growing trend in using protein dynamics
with heterogeneous interacting atoms, either as a metaphore or
as a mathematical representation, for understanding complex
biological organisms such as single cells and even tumor tissues
\cite{wolynes_sfi,fw_94, sasai_wolynes_03, Kauffman}.
Kinetic steady state is one of the
fundamental concepts in chemical and biochemical reaction
systems. Indeed, the notion of attractors has become increasingly relevant in studying cell differentiation and its fate determination,
and cancer non-genetic heterogeneity \cite{Kauffman, Gatenby, WangPNAS2011, huangsui, ao_hood, WangJRSI2014, WangCR2015, WangSR2016}.
Painted in a broad stroke, dynamics with dissipation usually
predicts a convergence of systems with different initial state.
How can such a picture be consistent, then, with an increasing
diversity and complexity as expected from living biological organisms according to Darwin's theory?

	In addition to a landscape perspective \cite{fw_91, WolynesScience1995, Graham, aoping,WangPNAS2008_1,WangADP2015, ge-qian-16} that
can be made very precise even for a large class of nonequilibrium systems
in terms of a generalized Gibbs function, complex
systems such as macromolecules, cells, and even biological
organisms also share other important characteristics.
The notion of {\em symmetry breaking} has been considered
by many thinkers as a fundamental mechanism for generating
complexity \cite{anderson,jjh,wolynes_96,
laughlin_wolynes,haken_00}.  At the core of this idea
are two elements: ($i$) a {\em singular point} in the phase
space of a nonlinear dynamical system where the future of
the dynamics is truly unpredictable \cite{prigogine_book}
and ($ii$) a {\em noise} ``too small to be taken into
account of by a finite being'' \cite{maxwell} with a lack
of detailed information for physical origin, or too
erratic to be fully comprehended by a rational person.

In modern mathematical theory of nonlinear dynamics,
($i$) is known as a ``saddle point'':
if a system is located precisely at the point and the dynamics
is absolutely deterministic, then it will remain there forever.
However, any infinitesimal perturbation will lead the system
away towards somewhere else. (Being precise, the
perturbation has to have a component along {\em unstable
manifold}. Physically, a perturbation exactly restricted on
a stable manifold is nearly impossible.) More importantly,
depending upon a particular perturbation,
there are in fact many possible outcomes,
or fates, which are seemingly chosen by chance.  In textbooks
illustrations, this is usually drawn as a
``double well potential'' with an ``energy barrier''
in between, see Fig. \ref{fig_02}, left pannel.  Chemists have
termed the saddle point a ``transition state'' \cite{eyring}.

What is the fundamental origin of ($ii$)?  For a macromolecule
immersed in an aqueous solution,
fluctuations are well understood in terms of atoms at finite
temperature \cite{rigler_book}:
They are too erratic to be meaningfully represented by
deterministic mathematics due to the lack of detailed, or
simply too much, information
of the motions of the individuals in the system.  Therefore, Fourier, Boltzmann, Einstein, Gibbs, Onsager, and
Kramers, together with many pioneers in the
physics of matter, have advanced a probabilistic description of
the states and dynamics of a complex, many-body system. (The probability in quantum physics has a fundamentally different origin.  Quantum dynamics is conservative while dynamics in thermal physics is dissipative. See \cite{leggett,mazur_bedeaux,esposito, ZhangJCP2014} for more discussions.)

The modern statistical theory of matter developed by physicists and chemists explains
the macroscopic, deterministic world in terms of the erratic,
stochastic dynamics of atoms and molecules.  Its fundamental
insight lies at the mathematical law of large numbers (LLN):  The
same law that gives Las Vegas casinos more confidence in their
profitability if more people are willing to gamble.  However,
another lesson from this story has not been told often enough:
An individual gambler with a nearly infinite amount of time and
money will win a jackpot, if the rule
of the game never changes.  Furthermore, there is actually
no logical causality between a person and a winning
event. (If anything, the logical causality
is in the mechanical movements of a slot machine; and
why and when a machine is picked due to physiological and
psychological processes \cite{jjh} of an individual gambler.)
Individual winning event is unpredictable.  What is
certain is that it will occur with probability 1 on any
given machine and the time to the occurrence is an
exponentially distributed {\em memoryless} random
variable \cite{taylor_karlin_book}.

    The successes of the LLN in classical
statistical mechanics have created an impression that
any system consisting of a large number of atoms and/or
molecules can be describable by deterministic mathematics:
Stochastic behaviors are averaged out. This
impression is rather misleading, especially
when dealing with nonlinear dynamics of a system consisting of
large number of individuals.   In fact, as we shall show,
stochasticity does not disappear in a wide class of systems that
have multistability, or the potential for phase transitions.
Here, stochasticities
simply manifest themselves as rare events on a
longer time scale; larger a system, longer the time.  Then
on an even longer time scale, numerous rare events constitute
another deterministic, continuous dynamics:  A single molecule
conformational transition is a rare event in Kramers'
theory, but they are the basis for the deterministic
kinetics of a chemical reaction system based on the
Law of Mass Action \cite{kramers,Wang_Wolynes1995,frauenfelder,qian_jsp}.

\section{Hierarchical Organization at Different Levels and Different Time Scales}

Indeed, P.W. Anderson stated that \cite{anderson} ``At each
level of complexity entirely new properties appear, and the
understanding of the new behaviors requires research which I think
is as fundamental in its nature as any other.'' He went on
to list a series of different levels of complexity: few-body
(elementary particle) physics, many-body physics, chemistry,
molecular biology, ..., physiology, psychology, and social
sciences. The elementary entities of each level in the hierarchy
obey the dynamic laws of a level lower, and at each level an
entirely new laws, concepts, and generalizations emerge, and
different treatment and theories are necessary.  Our mesoscopic
stochastic framework fits this hierarchy.
This hierarchy shares many features with the
organizational hierarchy among protein conformational
sub-states advanced by H. Frauenfelder and coworkers
\cite{fw_91,fw_94,frauenfelder}:
Going downward, a protein consists of
secondary structural motifs, which consists of amino acids,
which consists of atoms, etc.  Going upward, a cell consists
of a large number of macromolecules, and a tissue consists
of a large population of cells, etc. As an example, a recent study used contact geometry ideas to examine how to move complex descriptions of a system from one level to another \cite{Grmela}. In our framework, we expect coarse-grained descriptions emerge from the collective dynamics.

     The chemical reaction theory, together with protein
science, serve as a paradigm for stochastic, mesoscopic complexity
\cite{wolynes_sfi,fw_94,WangADP2015,laughlin_wolynes,frauenfelder,qian_jsp,qian_part_1}. Consider a chemical transformation $A+B\rightarrow C$
in an aqueous solution.  The conceptual framework for
such a reaction developed by Kramers \cite{kramers}
is now accepted as the theoretical foundation of chemical
reactions in condensed phase. In fact,
a chemical reaction theory involves {\em three}
levels in Anderson's hierarchy: ($a$) few-body physics
detailing the collisions of a few water molecules (H$_2$O)
with a few atoms within reacting molecules; ($b$) many-body
physics concerning molecules $A$, $B$ and $C$ in a sea of solvent
molecules; and ($c$) chemistry whose elementary events are
discrete chemical reactions such as $A+B\rightarrow C$.  In
($a$) one is concerned with collisions leading to high-frequency
vibrations; ($b$) is primarily interested in the {\em mechanism}
and process
of how $A$ and $B$ collide, interact and formation of $C$ occurs in
terms of the atoms in the molecules with vibrations and diffusion,
while in ($c$) the actual chemical reaction is represented by a
single second-order rate constant for a discrete transition.  Fig. \ref{fig_01}A uses a schematics to illustrate this hierachy.

    Kramers' theory predicts that these different
levels also translate to different time
scales \cite{wolynes_sfi}: At the time scale of a molecular
reaction, the process in ($a$)  is so fast that it can be essentially treated
as infinitely rapid fluctuating dynamics with certain appropriate
statistics.  The time scale for processes in ($b$) is of course determined by
the energy and force in the molecular system which Kramers called
``a field of force'', and the outcome of the theory is a discrete
event of an elementary chemical reaction whose time scale, $\sim 10^{-7}$
sec., is almost infinite on the time scale of ($a$), $\sim
10^{-12}$ sec.  Kramers' mathematical theory is one of the first
that reveals an interplay between``chance'' and
``necessity'' \cite{monod_book,haken_book,ao_plr, WangJCP2012}.

    Now for a cellular biochemist who is interested in
a metabolic system with many biochemical species, the individual
$A+B\rightarrow C$ on the sub-$\mu$sec scale is just
part of fluctuations.  He/she is interested in the dynamics
of how various concentrations of metabolites change with
time.  By using Law of Mass Action, steady states
of the biochemical reaction system which can be reached
on the order of seconds, can be predicted.
Then on an even longer time scale, recent studies on phenotypic
switching point to the stochastic transitions from
one biochemical steady state to another in a single
cell, on the time scale of $10^3$ sec.
Again, to $10^{1}$, $10^3$ is essentially infinite.
Figs. \ref{fig_01}B again shows an illustrative schematics.

    One of the deepest concepts developed by chemists
in connection to chemical reactions is the notion of ``transition
state'' \cite{eyring, WangCS2014}. We see that it is at the very transition
state the dynamic of a symmetry breaking occurs
in molecular systems. If a molecular system is infinitely
large, then this
symmetry breaking is static: the chance of Kramers' barrier
crossing could take the time as long as the age of the universe.
This is the symmetry breaking picture of Anderson \cite{anderson}.
However, a macromolecule such as a protein can in fact
jump among its different {\em conformational states} on the time
scale observable in a laboratory \cite{fw_94,rigler_book,frauenfelder}
and exhibits successive
{\em dynamic symmetry breaking} \cite{jjh}. In this latter case, a
discrete-state stochastic description of the dynamics in term of a
Markov jump process is most appropriate \cite{ge-qian-16,ge_dill,qian_iop}. As
differential equation approach to classical dynamics, Markov
approach to stochastic dynamics is very general. Even certain
processes with long memory can be mathematically transformed into
a Markovian representation.

A unavoidable mathematical issue is at the heart of
any theory of mesoscopic systems.  As Anderson pointed
out in \cite{anderson} that ``It is
only as the nucleus is considered to be a
many-body system | in what is often
called the $N\rightarrow\infty$ limit | that such
emergent behavior is rigorous definable.''  The
importance of taking thermodynamic limit in a mathematical
theory of phase transition goes back to
Kramers in 1936 \cite{kramers_bio}.
The thermodynamic limit according to textbooks on
equilibrium physics of matters is to
take the time $t\rightarrow\infty$ first
and then systems size $N\rightarrow\infty$ afterward.
In fact, the $t\rightarrow\infty$ limit never appeared
since {\em equilibrium} is assumed at the onset.
On the other hand, for nonlinear dynamic
behavior in a macroscopic system, one often takes the
$N\rightarrow\infty$ limit first for finite $t$.  In fact,
nonlinear, emergent dynamic behavior of a complex
system can {\em only} be rigorously defined with $N\rightarrow\infty$
\cite{ge-qian-16}, followed with
fluctuation analysis with finite $N$.
Our framework is explicitely concerned
with the order of these two limits \cite{ZhengWM}.
In reality, both limits
are simple idealizations.  Still, each limiting procedure
has it validity on an appropriate time scale: relaxations
within a basin of attraction and inter-attractor
stochastic jumps.

\section{Nonlinear bistability,
bifurcation, and phase transition}

    In the mathematical theory of deterministic nonlinear dynamics,
symmetry breaking is intimately related to the problem of
{\em saddle-node bifurcation} \cite{strogatz_book}.  In fact, the
theory of saddle-node bifurcation and its topological
representation, known as {\em catastrophe theory}, is exactly a
symmetry breaking problem viewed in a relatively short time scale.

This section should be followed closely
with Fig. \ref{fig_03} at side.  It is essentially a
mathematics exercise, with an explicit example given in
the Appendix.  For broader audiences, however, we
choose to carry out the discussion using verbal
narratives as much as possible.
Let us consider an ordinary differential equation (ODE)
for 1-d $x(t)$, $\rd x/\rd t = b(x;\alpha,\beta)$ with two
parameters $\alpha$ and $\beta$.
In other words, for each pair of $(\alpha,\beta)$, there
is an ODE.  Let us further assume that for some values of
$(\alpha,\beta)$ the ODE has only a single stable steady
state (fixed point), and for other parameter values there
are three steady states, two stable $x_1^*$ and $x_2^*$,
and one unstable $x^*_3$ in between: $x_1^*<x_3^*<x_2^*$.
Note the $x^*$s are functions of $\alpha$ and $\beta$; in
fact, all the $x^*$s irrespective of $1,2$, or $3$ are roots
of $b(x,\alpha,\beta)=0$.  This argebraic equation defines
a surface shown in Fig. \ref{fig_02}B.  When a piece of
paper is gently folded, the multiple steady states $x^*$
as functions of $\alpha$ and $\beta$ form a multi-layer
surface in 3-d.

    In classical van der Waals gas problem, the $x$ variable is
equilibrium volume of a box of gas, $\alpha$ and $\beta$ are temperature and pressure.  In a biochemical phosphorylation
feedback system, $x$ is the fraction of phosphorylated
protein, $\alpha$ and $\beta$ are the kinase activity and
ATP phosphorylation potential \cite{ge_qian_prl,ge_qian_2}.
The equation $b(x;\alpha,\beta)=0$ is known
as an ``equation of state'' in van der Waals theory,
and an ``equation of phosphorylation-dephosphorylation
switch'' for signaling module \cite{beard_qian_book} .

    In Fig. \ref{fig_02}B, projecting
the three layers to the $\alpha\beta$ plane for the two
parameters, topologist Ren\'{e} Thom had the deep insight
that the region has to have a wedged shape with a cusp as
shown in Fig. \ref{fig_02}B, also in
Fig. \ref{fig_03}A \cite{thom_book}.  Now if you
keep $\beta$ constant and vary $\alpha$ acorss the
wedged region starting from far left, as illustrated
by the dashed blue line in Fig. \ref{fig_03}A, the number
of steady states changes from 1, to 2 to 3, and
back to 2 and 1.  This is shown in Fig. \ref{fig_03}B.
The black $S$-shaped curve is a ``bifurcation
diagram'' which shows the position of steady state(s)
as a smooth function of $\alpha$ (with a given $\beta$).

    At the blue vertical lines in  Fig. \ref{fig_03}B,
the phenomena of changing number of steady states
are called {\em saddle-node} bifurcation.  For
small and large values of $\alpha$, the system
has only a single steady state (fixed point).
The blue dashed lines mark the critical $\alpha$ values
at which there is a sudden appearance or disappearance
of a pair of stable and unstable steady states.
The pair ``bursts out of blue''; thus acquired the
name of blue sky bifurcation \cite{strogatz_book}.
One of the extensively studied examples of this type of
behavior in biochemistry is forced molecular dissociation
leading to non-covalent bond rupture \cite{shapiro_qian}.

    So far, we have discribed how bifurcation
arises in nonlinear, deterministic systems with bistability.
A deterministic nonlinear approach is usually valid for
macroscopic systems. From a mesoscopic perspective, this
means all our above discussion starts with a system without
fluctuation.  More precisely, when one studies a mesoscopic
molecular system, the numbers of individuals of various
species in a system $\vec{N}=(n_1,n_2\cdots)$ and the volume
$V$ of the system, are usually specified.  The ODE perspective
is for infinitely large system, i.e., introducing
``concentration'' $\vec{x}(t)=\vec{N}(t)/V$, and then
mathematically taking $\vec{N},V\rightarrow\infty$ to obtain a
``macroscopic limiting behabior'' in terms of the nonlinear
dynamics for $x(t)$.  Then in the limit of
$t\rightarrow\infty$, multiple attractors is revealed.
Different initial conditions will leads to different
steady states. Because of this procedure, the transition
between two fixed points, the most important consequence
of fluctuations, is not possible in the deterministic
analysis.  There is a breakdown of ergodicity.

    This is not the {\em thermodynamic limit} which requires
a true equilibration among all different attractors.
In fact, the most important information missing in the
ODE analysis is the relative weights for different
attractors.  This comes from analysis based on probability
and stochastic processes. A finite-size correction naturally
introduces stochasticity.  Depending upon the chosen
representation for a system, a finite-size mesoscopic
model can be a discrete or continuous stochastic process.
For example, the dynamic equation for a stochastic
concentration $x(t)$ can be characterized by
$\rd x(t) = b(x)\rd t + V^{-\frac{1}{2}}\rd B_t$ where
$V$ represents the size of the dynamical system and $B_t$ represents
a Brownian motion fluctuation.  One notices that if $V\rightarrow\infty$,
the dynamics is reduced to the above ``macroscopic limit''.

    To study the true thermodynamic limit, one lets $t\rightarrow\infty$ first in a stochastic model
followed by $V\rightarrow\infty$.  This way, an initial
value independent (i.e., ergodic) probability distribution
across all attractors emerges.  The mathematical theory
for this type of stochastic differential equation
(i.e., nonlinear Langevin equation) shows that the
stationary density has the form \cite{ge_qian_ijmpb}:
\begin{equation}
     f^{st}_V(x) = \mathcal{C}_V e^{-V\phi(x)},
\label{eq_1}
\end{equation}
in which $\mathcal{C}_V$ is a normalizationm factor.
Furthermore, $\phi(x)$ has local minima at $x_1^*$
and $x^*_3$ and a maximum at $x_3^*$.  This means the
probability distribution $f_V^{st}(x)$ peaks at
$x_1^*$ and $x_2^*$.  It it the extrema of function
$\phi(x)$ that match the fixed points of $b(x)$.
The behavior of the deterministic dynamics is closely
related to the {\em modal values} of the finite,
mesoscopic system rather the {\em mean value}
\cite{qian_arbp}.

The shapes of $\phi(x)$, the ``landscapes'' \cite{WangJRSI2014, aoping, WangPNAS2008_1, WangBJ2007, WangPLCB2007, WangJCP2010, WangPNAS2014, qian_ge_mcb}, for different $\alpha$ and $\beta$ are shown in
Fig. \ref{fig_03}A and B.  For each $(\alpha,\beta)$
there is a $\phi(x;\alpha,\beta)$.  Along the dashed
blue line in Fig. \ref{fig_03}A, the corresponding
$\phi(x)$s are illustrated in orange, as well as
the black shapes shown below the curve in \ref{fig_03}B.
When changing $\alpha$ horizontally in \ref{fig_03}A,
the landscape for bistability develops a bias for
one of the minima.  Outside the wedged region,
one of the minima disappears all together.

    Note that Eq. \ref{eq_1} is obtained by letting
$t\rightarrow\infty$ while keeping $V$ finite.
Now different phenomenon arises if one lets
$V\rightarrow\infty$ in (\ref{eq_1}): The distribution
$f_{\infty}^{st}(x)$ will concentrate at
the {\em global minimum} of $\phi(x)$ with probability 1!
Even though a system can be bistable or multistable,
a metastable state, i.e., the non-global local minimum of
$\phi(x)$, has zero probability in the limit of $V\rightarrow\infty$, if one allows the system to truly
equilibrate.

    Therefore as predicted by the LLN, generically there
is a {\em unique} steady state for a bistable (or multistable)
system in the true thermodynamic limit, except at a critical $\alpha^*$ when the two minima are precisely equal
$\phi(x_1^*,\alpha^*)=\phi(x_2^*,\alpha^*)$, shown in
Fig. \ref{fig_02}B, and the dashed red curve in
Fig. \ref{fig_02}A.(It is also clear
that if the correlation in such a system at
$\alpha^*$ were short-ranged, then there would be infinite number of ``identical, independent subsystem'', which would imply
LLN being again valid.  Thus, the violation of LLN at $\alpha^*$
dictates an infinitely long, non-exponential decay correlation
in the system.)  For each given $\beta$ value, there
exists an $\alpha^*(\beta)$.

    Therefore, in the limit of $t\rightarrow\infty$ followed
by $V\rightarrow\infty$, i.e., in the true thermodynamic limit,
the $S$-shaped curve in Fig. \ref{fig_03}B is no more;
only a discontinuous jump at $\alpha^*$, marked by the
red dashed vertical line.  This is a reminiscent of the
Maxwell construction \cite{nicolis_lefever} for the
van der Waals theory of non-ideal gas.  Similarly,
in Fig. \ref{fig_03}A the wedge is no longer relevant;
only the dashed red curve which should abruptly terminates at
the cusp.   This is known as a {\em first-order phase
transition line}.

    Now let us focus on the cusp in Fig. \ref{fig_03}A.
Moving along the red curve inside the wedged region, the
landscape changes from symmetric bistable with two
equal minima to a monostable single minimum, when
passing the cusp.  This is magnified in Fig. \ref{fig_03}C.
It is L.D. Landau's {\em second-order phase transition} \cite{landau_book}.  In nonlinear dynamical systems theory,
crossing the cusp constitutes a robust pitchfork bifurcation \cite{strogatz_book}.

    Therefore, a kind of symmetry and symmetry breaking
emerge in the simple bistable nonlinear system with
stochasticity, i.e., a complex mesoscopic system. Table
1 summaries the discussions above: In the left column,
$V\rightarrow\infty$ with $x=N/V$, followed by
$t\rightarrow\infty$; in the middle column
$t\rightarrow\infty$ while holding $V$ finite, and
the right column gives the true thermodynamic limit
as $t\rightarrow\infty$ followed by $V\rightarrow\infty$.

    How is this ``mathematical'' description of bifurcation
and phase transition related to the traditional physics
of matters?  Thermodynamic description of a system
neglects all time scales that are too long to be
of interests, then assumes that all the remaining time scales
reach an ergodic stationarity. We emphasize this
point since strictly speaking there are always slow
processes in reality.  For example, there is a slow rate
of peptide hydrolysis in an aqueous solution for any
protein molecule. But this effect is usually neglected,
rightly, in the thermodynamic theory of protein conformational
transitions.  Also,
according to Newton's third law, there is always
a {\em consequence} at the origin that generates a force.  For
example a magnet that induces supercurrent in a
type II superconductor decays slowly.
Deterministic nonlinear dynamic description
of a system, on the other hand, considers interesting
dynamics in the medium time range: Different initial
conditions will lead to different steady states.  The
true equilibration among different steady states,
however, is out of reach in a deterministic description.
A mesoscopic system can exhibit rich behavior precisely
because both scenarios are accessible and they even
interact \cite{WangADP2015, qian_part_1, haken_book, nicolis_lefever}.

    A {\em mean-field} treatment in the classical physics
usually entails deriving a relation among ``mean values''
by neglecting fluctuations.  Therefore, it corresponds to $N,V\rightarrow\infty$ first.  Thus often it is incapable
of reaching the ergodic thermodynamic limit.
Furthermore, the cusp is precisely the critical point
$T_c$ in Landau's phenomenological phase transition of
ferromagnet in terms of free energy
$F(x)= (T-T_c) x^2 + b x^4 - Jx$ with $J=0$ \cite{gaite}.
This is the symmetry breaking picture generally
discussed in the theory of phase transition.
Changing $\alpha$ and $\beta$ correspond to changing
$J$ and $T$ in the above $F(x)$.  The essence
of Landau's theory is a bistable system with
stochasticity (noise) \cite{haken_book}.

   Yang and Lee have established a general mathematical
origin for phase transition \cite{yang_lee,lee_yang}.
They have shown that the mathematical non-analyticity, a necessary
feature of any rigorous phase transition theory,
is related to a zero of a partition function moving
from complex plane onto real axis in the $V\rightarrow\infty$
limit (Recall that the free energy is
the logarithm of a partition function.)  It has been demonstrated
recently that this same mathematical description applies to any
bistable system with stochastic elements
\cite{ge_qian_prl,ge_qian_2}, including mesoscopic biochemical
system with bistability.   Therefore, the notion of
phase transition, together with key concepts such as
symmetry breaking and the perspective of ``true
thermodynamic limit'' have a broad applicability to
systems exhibiting phenomena as catastrophe,
rupture, and hysteresis \cite{shapiro_qian}.  It is a
complementary description of bistability in the presence
of stochasticity.

The notion of symmetry breaking used in \cite{anderson,jjh}
is intimately related to bifurcation. While
bifurcation of a ground state in solid-state and particle
physics is due to a symmetry in a Hamiltonian \cite{fw_94},
nonlinear complex systems have symmetries, with
canonical forms, at their bifurcation
points \cite{strogatz_book}.  However, the existence
of multistability and attractors is often an emergent
phenomenon itself. Our above discussions
illustrate that with stochasticity,
bifurcations in the true thermodynamic limit exhibit
phase transitions | a probability distribution has
certain symmetry; a realization by a particular system
breaks the symmetry.  In fact the pitchfork bifurcation is
a necessary part of a catastrophe phenomenon.

\section{Emergent discrete stochastic dynamics in
nonlinear systems with multiple attractors}

    The forgoing discussion focused on systems with
bistability.  For a highly nonlinear, complex system with
stochastic elements, there could be a large number of
attractors.  Therefore, on a time scale much longer than the
deterministic dynamics that occurs within each basin of
attraction, inter-attractor dynamics can be represented as a Markov
jump process among a set of discrete states. This is an insight
one learns from macromolecular dynamics like those of a protein: Kramers'
theory accounts the transition rates, usually with an exponentially
distributed waiting time, between each pair of
``conformational states''; but the dynamics of an enzyme is
usually represented by chemical kinetics which represents the
conformational states in discrete terms. More interestingly, in
recent years the Delbr\"{u}ck-Gillespie approach to nonlinear
biochemical reaction systems treats each elementary chemical reaction in a
mesoscopic volume as a stochastic process, and derives endogenous
phenotypic ``cellular states'' \cite{sasai_wolynes_03, ao_hood, WangPNAS2008_1, WangADP2015, WangBJ2007, WangPLCB2007, WangPNAS2014}
and cellular evolutionary dynamics \cite{WangPNAS2011,qian_iop}. Robustness of a cellular state and punctuated equilibrium in state transitions are necessary consequences of this dynamics description.

    The time scales, short or long, are defined by the Kramers'
theory for a barrier-crossing process.  It is tempting to suggest
that complexity at a mesoscopic scale originates from a
system with the size at which both the deterministic,
converging dynamics on a short-time and the stochastic,
diverging dynamics on a long-time are accessible to an
observer \cite{fw_94,laughlin_wolynes}.
We stress that the stochastic
fluctuations on the very fast time scale yield the
divergent behavior of stochastic jumps out of the basins
of attraction on a very long time scale \cite{fw_94}.
Conversely, interactions between dynamics at these different time scales
lead to slow dynamics modulating the fast motions with possible eddy current \cite{sasai_wolynes_03, WangSR2016, WangADP2015, WolynesPNAS2005, HornosPRE2005, WangJPCB2011, FengBJ2012, WangPNAS2013}.
A stochastic system in general moves toward states with
higher probability, exhibiting a form of contingency \cite{fw_94}, or adaptation \cite{tu_1,tu_2}. In enzyme kinetics, this is the origin of dynamic disorder
\cite{frauenfelder}. It is also illuminating to
point out that an {\em ab inito} computation of the emergent
stochastic transitions from the detailed dynamics can range from a
highly challenging task to practically infeasible
\cite{laughlin_pine}: In protein folding, this is only
accomplished very recently for a single transition of biochemical
significance based on atomic-level molecular dynamics \cite{shaw}.

\begin{figure}[ht]
\[
\includegraphics[width=3.5in]{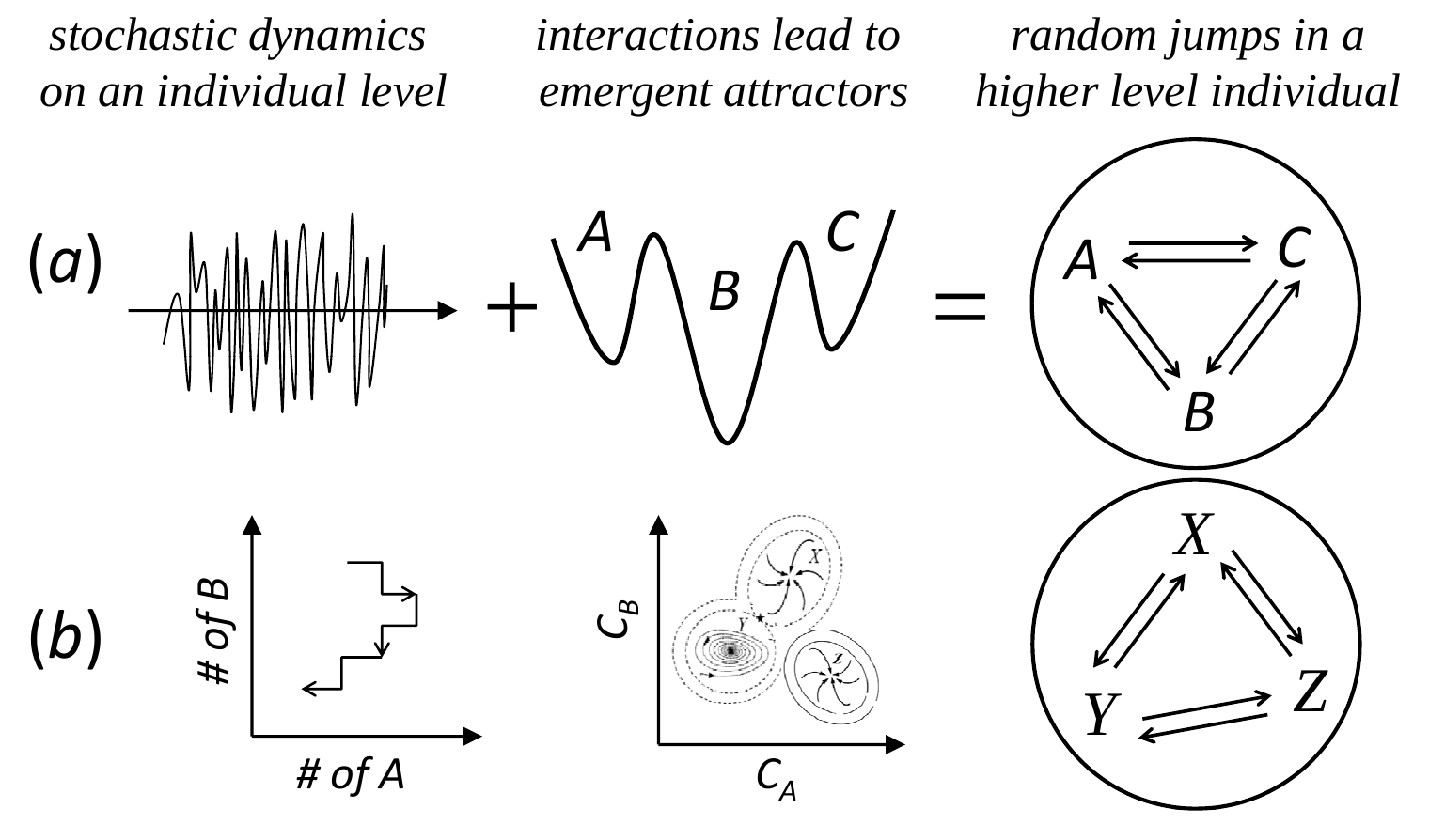}
\]
\caption{($a$) A schematics showing rapid solvent-macromolecule
collisions, as a source of stochasticity and together with
a multi-energy-well landscape, gives rise to a kinetic jump
process for an individual macromolecule with multiple
states (shown within the circle). ($b$) A level higher,
many interacting chemical individuals each with multiple
discrete states form mesoscopic nonlinear reaction systems.
In the space of copy numbers and concentrations of
chemical species, such a system exhibits Monte carlo walk
due to each and every stochastic reaction with emergent
multiple nonlinear attractors ($C_A$, $C_B$ represent
concentrations of $A$ and $B$). The stochastic transitions
within each macromolecule serve as ``noise'' leading to
state switching on the whole reaction system level, shown
by $X$, $Y$, and $Z$.  ($a$ \& $b$) The same schematics
illustrates the Anderson's hierarchy of complexity:
There is a randomness in the dynamics of an "individual", be
it a macromolecule in aqueous solution, a cell in a
tissue, a trading company in an economic system, or an animal in
an ecological environment. Interactions between individuals in a
system form a nonlinear dynamical system with emergent
attractors. The fundamental insight of Kramers' theory
is that, while the Law of Large Number is at work, there
will be emergent stochastic dynamics, beyond the infinite
time of a deterministic nonlinear dynamics, at the systems
level in an evolutionary time scale. This
randomness, represented by the individual jumps inside the
"particle" on the right, then becomes the stochastic element
for the nonlinear dynamics of an organism, a level higher,
that consists of many such interacting particles.}
\label{fig_01}
\end{figure}

    Let us reiterate: The insights from the present
discussion which departs from the LLN perspective is that
stochasticity does not completely disappear in a reasonably
large, nonlinear mesoscopic system.
Rather, it manifests itself
as a stochastic jump process on a much longer
time scale among a set of discrete states, as has been
shown in Fig. \ref{fig_01}A and B.  These discrete states
are attractors of an interacting nonlinear dynamical system.
These discrete states are determined by the behavior and interactions
among individuals within the system; yet their existence,
locations, and transition times are completely non-trivial.
They are emergent properties.  Well-known examples are
cooperativity in equilibrium physics
and feedbacks in biological networks \cite{qian_arbp}.
In the former, cooperativity leads to crystallization; and
in the latter nonequilibrium systems, feedbacks
lead to self-organization. The nature of emergent
phenomena is a consequence of nonlinear interactions between
individuals \cite{prigogine_book,haken_book,laughlin_pine}.

    The emergent stochastic jump dynamics of an individual,
of course, becomes the stochastic elements for a higher level
population system in a larger space
with longer time. In this fashion, Anderson's hierarchy moves up
level by level with the nonlinear, stochastic dynamic
framework \cite{qian_part_1}.  This is again illustrated in Fig. \ref{fig_01}.
Recall the example of a single biological cell as a mesoscopic
chemical reaction system where this hierachy has already been
appreciated \cite{wolynes_sfi}: Boltzmann started the tradition of
treating molecular collisions as stochastic elements in a kinetic
theory. Kramers has shown that the discrete chemical reaction can
be described as Brownian motion in a force field
\cite{kramers,hanggi}. Supported by the recent experimental
advances in single-molecule chemistry and biophysics
\cite{rigler_book}, Delbr\"{u}ck-Gillespie's theoretical approach
to chemical reactions considers each chemical reaction as a
stochastic jump, and derives ``states'' and ``dynamics'' of
cell-size biochemical reaction {\em systems} \cite{WangPNAS2011,
WangPNAS2008_1, WangADP2015, qian_iop, WangJCP2010,WangPNAS2014}.

Continuing with this perspective, the question of whether
epigenetic phenotypic states at a cellular level is a part of intrinsic biochemical dynamics or an external, environmentally
induced phenomenon can be addressed using a mesoscopic dynamic
approach \cite{WangPNAS2011, ao_hood,WangPNAS2008_1,WangADP2015, qian_arbp,WangBJ2007}.
Epigenetic switching could indeed be viewed as a
``phase transition'' in mesoscopic biochemical
systems \cite{ WangADP2015,ge_qian_prl,ge_qian_2,qian_ge_mcb}.
Phase transition that might play a possible role in the emergence of
life itself, as suggested recently in \cite{cronin-walker}, should be
understood as such.

\section{Conclusions}

The mathematical concept of thermodynamic limit is defined
as an infinitely large system reaches its infinitely
long-time limit, where the long-time has to be sufficient
to overcome all the exponentially small barrier crossing
probabilities.  Therefore, it is immediately clear that
there are two competing limits for time and systems
size, and the order of taking these limits matters.  Complex
behavior arises when these two limits are not exchangeable
due to {\em non-uniform} convergence \cite{vellela_qian_09}.
As we have discussed, a real thermodynamic
limit, which takes $t\rightarrow\infty$ first,
is simple.  A mesoscopic system is messy | When
Anderson talked about $N\rightarrow\infty$ \cite{anderson},
there are in fact two possibilities: finite time
dynamics requires taking it before $t\rightarrow\infty$,
and thermodynamics requires taking it after
$t\rightarrow\infty$: The latter produces a simpler picture of
the world with universality, and the former produces a much
more complex picture of a world with diversity.

The present discussion focuses on the
emergence of discrete transitions between different
dynamic basins.  While we have not explicitly
discussed system with spatial characteristics, we
believe a large part of the discussions is applicable
to stochastic reaction-diffusion systems \cite{murray_book, WuJPCB2013, WuJCP2014_1, WuJCP2014_2}.
Recent work also points to the important phenomena
associated with time symmetry breaking in nonequilibrium
steady state of mesoscopic systems \cite{WangPNAS2008_1,WangADP2015, WangJCP2012, WangJCP2010, zqq,seifert,WangBJ2010, WangJCP2011}.  A
deeper understanding toward the relationships among
different forms of symmetry breaking in space, time,
and dynamics remain to be
elucidated \cite{wolynes_96,yangcn}.

In physics, the notion of mesoscopics often refers to dynamics
such as conductance fluctuations in small size devices. In the
present work we see the scope of ``mesocopic phenomena'' to be
much broader: It can also cover many other interesting
behavior including biochemical cells with self-organizations
\cite{qian_iop, zqq}.  In fact, it is the description in terms of
stochastic nonlinear dynamics, incorporating both chance and
necessity \cite{monod_book}, that gives the ``middle way''
\cite{laughlin_wolynes} a
unique yet universal characteristics \cite{ aoping, WangPNAS2008_1, WangADP2015, haken_book, WangJCP2012, qian_iop,WangJCP2010, WangJCP2011}.
This is one of the most fundamental insights of J.W. Gibbs, and
the contribution of chemical science to the theory of complexity
\cite{wolynes_sfi}.

\section*{Acknowledgments}
The authors thank S. A. Kauffman, A. J. Leggett, and P. G. Wolynes
for helpful discussions, and NSF and NSFC for
financial supports.

\begin{center}
Table 1: Terminologies and Phenomena in Infinite-time Dynamics$^1$
\end{center}

\begin{tabular}{|c|c|c|}\hline
Deterministic$^2$ &\multicolumn{2}{|c|}{Stochastic$^{3,4}$}\\
\hline
{} &$V<\infty$ &$V=\infty$ \\ \cline{2-3}
basin of attraction  &landscape well  &\\
{} & &global landscape minimum,\\
stable fixed points &landscape minima &all local minima have zero\\
(attractors) &{} &probability \\
unstable fixed point &landscape maxima & \\
(repellers) &{} & \\
{} &{} &{}\\
out-of-blue saddle- & emergence of a pair of &n/a\\
node bifurcation &local min $\&$ max &{} \\
{} &{} &{}\\
bi-stability wedged & double-well region &wedged region collapses\\
region &{} &into a coexistence line,\\
{} &Maxwell construction &$1^{st}$ order phase transition\\
n/a &for two equal minima &{}\\
{} &{} &{}\\
cusp &two equal minima &critical point\\
{} &at the boundary of &{}\\
pitchfork bifurcation &the wedged region &$2^{nd}$ order phase transition\\
{} &{} &{}\\
\hline
\end{tabular}

$^1$ n/a means a phenomenon has no correspondence,
and significance, in this setting.

\noindent
$^2$ Deterministic means one takes $V\rightarrow\infty$
first to obtained an ODE for $x=N/V$, i.e.,
macroscopic limit, followed by $t\rightarrow\infty$ to
obtain steady states (attractors) of the determinstic
nonlinear dynamics, starting with different initial
values.

\noindent
$^3$ Stochastic dynamic with very large but finite
size ($V<\infty$) has a proper probability density for
its stationary process (i.e., $t\rightarrow\infty$
while holding $V$):
$f^{st}_V(x)=\mathcal{C}_Ve^{-V\phi(x)}$ where $\phi(x)$ is a landscape.

\noindent
$^4$ A true thermodynamic limit takes the results in the
middle column, followed by $V\rightarrow\infty$.
Because $f_V^{st}(x)$ is normalized,
the limit $f^{st}_{\infty}(x)$ has a singular
support, with probability 1 concentrated on the global
minimum of $\phi(x)$.

\begin{figure}[ht]
\[
\includegraphics[width=3.0in]{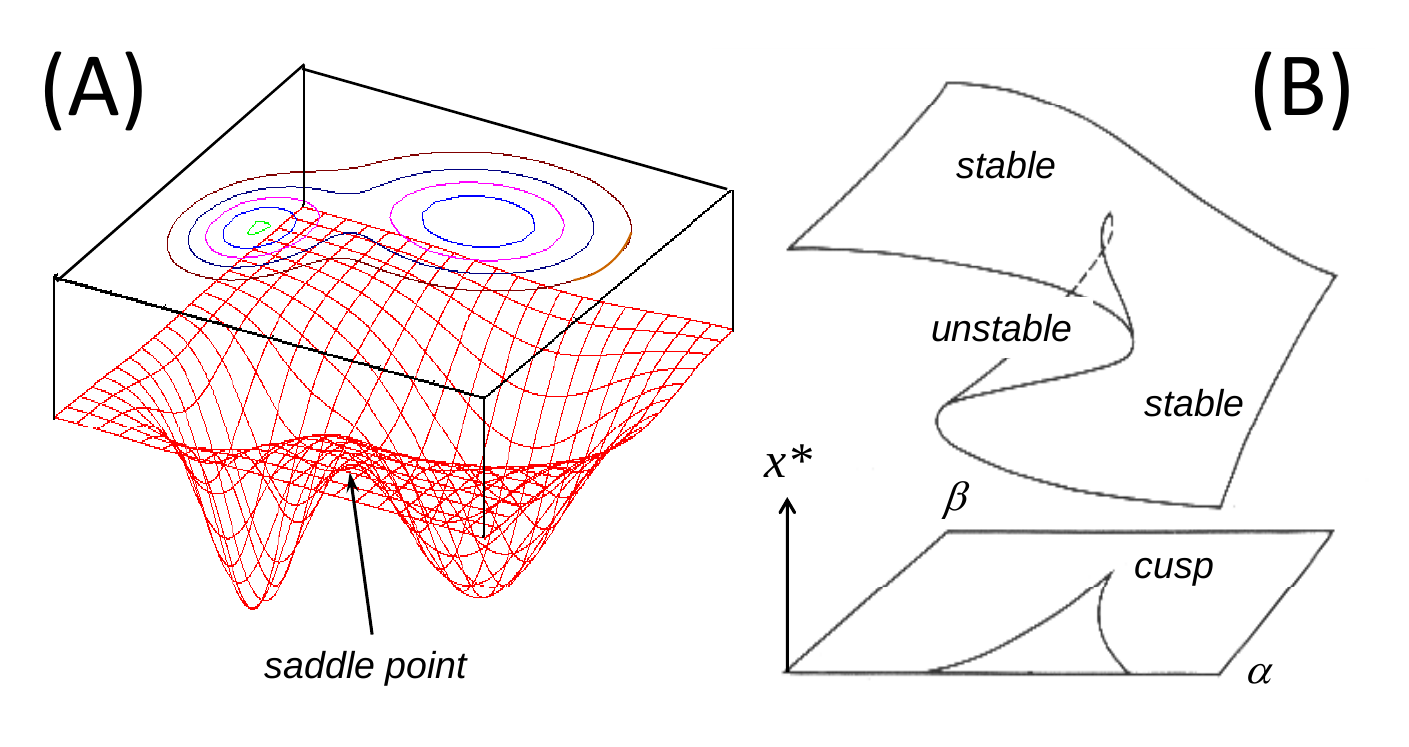}
\]
\caption{Double wells with a saddle point
and three-layer fold with a cusp.  (A) The surface
represents an ``potential function'' and the contours
are iso-energy lines.  Moving from one well to another,
the most likely path with the lowest ``barrier'' to overcome
is through the saddle point, i.e., a mountain pass.
(B) The fixed points of an ODE $dx/dt=b(x;\alpha,\beta)$ are
the roots of the equation $b(x;\alpha,\beta)=0$.  $x^*$ as
a function of $\alpha$ and $\beta$ is a surface in 3-d.
Folding a smooth surface into three layers, there
is necessarily a cusp in the $\alpha\beta$ plane.}
\label{fig_02}
\end{figure}

\begin{figure}[ht]
\[
\includegraphics[width=2.55in]{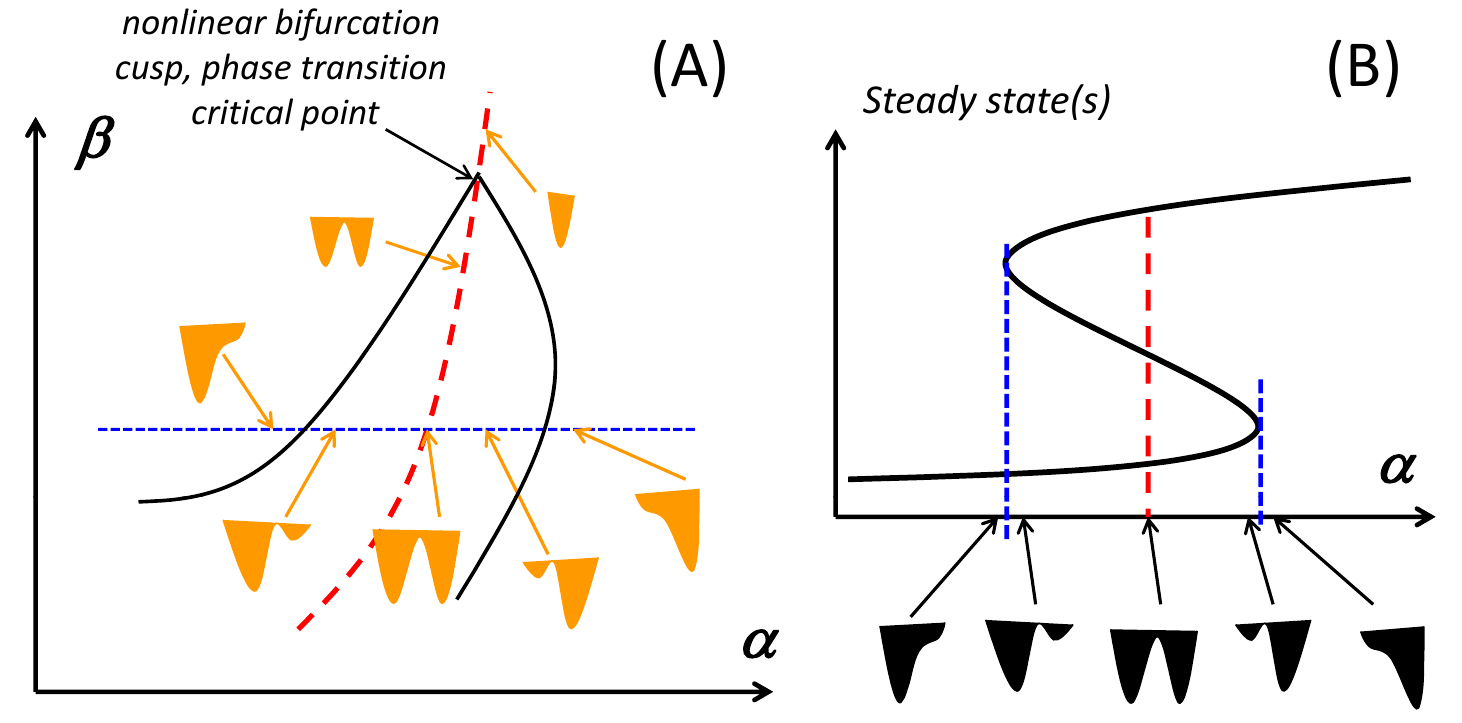}
\includegraphics[width=0.85in]{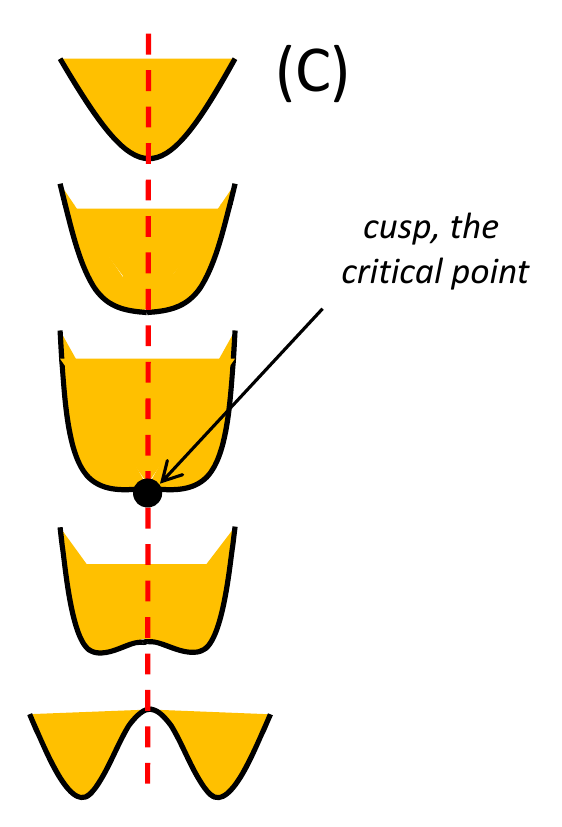}
\]
\caption{For a mesoscopic model, if one takes its size,
i.e., $N,V$ to $\infty$ first before $t\rightarrow\infty$,
as usually done in a mean-field treatment, one obtains
a nonlinear dynamics with broken ergodicity.  The
infinite-time limit of such a deterministic system can be
described by black curves in (A): there is the
well-known cusp catastrophe represented by a wedged region,
as the projection in Fig. \ref{fig_02}B.  Inside (outside)
the wedged region, the dynamical system has three (one)
fixed points.  For a fixed value of $\beta$ while varying
$\alpha$, the horizontal blue dashed line crossing the
boundary of the wedged region correspond to the two vertical
blue dashed lines in (B), where two two saddle-node bifurcations
occur.  For a true thermodynamic limit, however, one first
takes $t\rightarrow\infty$ for large but finite $N,V$ and
obtains a system's stationary probability distribution
$f_V^{st}(x)=\mathcal{C}_Ve^{-V\phi(x)}$, where $x=N/V$ and $\phi(x)$ function is the dynamic landscape.  $\phi_V(x)$ is
shown as the orange and black shapes in (A) and (B).  Then
one lets $N,V\rightarrow\infty$ and obtains $f_V^{st}(x)\rightarrow
\delta(x-x^*)$ where $x^*$ is the global minimum of the
$\phi(x)$. Marking the co-existence of two equal minima,
a Maxwell-like construction shown by the verticle red
dashed line in (B) is introduced: a discontinuity
appears in the $x^*$ as a function of $\alpha$,
at $\alpha^*$.  In (A), the red dashed line represents
the line of $\alpha^*$ for different $\beta$.
The cusp of the wedged region matches the critical
point in the phase transition theory.  In fact, as
magnified in (C), along the red phase
transition line in (A), there is a pitch-fork bifurcation at
the critical cusp.  In thermodynamic limit this is known
as second-order phase transition. }
\label{fig_03}
\end{figure}

\nocite{*}


\appendix

\section{A mathematical example}

    The schematics in Fig. \ref{fig_03} can be
mathematically produced through the following
example of a cubic system.  It is a simple
variation of L.D. Laudau's energy function
$F(x)= (T-T_c) x^2 + b x^4 - Jx$.

{\bf\em A. Ordinary differential equation with bistability.}
Ordinary differential equation (ODE)
\begin{equation}
      \frac{dx(t)}{dt} = -\left(x^3+x^2
                -\alpha x+\beta\right) = b(x)
\end{equation}
has fixed point(s), or steady state(s), as the roots of cubic
polynomial
\begin{equation}
           b(x)= -\left(x^3+x^2-\alpha x+\beta\right)
               = 0.
\label{beq0}
\end{equation}
Its solutions $x^{ss}(\alpha,\beta)$, as a multi-layer surface
with fold, is a function of $\alpha$ and $\beta$.  In the
$(\alpha,\beta)$ plane, the region in which $x^{ss}(\alpha,\beta)$
takes three values has wedged shape with a cusp (e.g., Fig.
\ref{fig_02}B). The boundary of the wedged region is given by a
parametric equation with $-\infty<\xi<+\infty$:
\begin{equation}
 \left\{  \begin{array}{ccc}
             \alpha(\xi) &=&  3\xi^2+2\xi \\
             \beta(\xi) &=&  2\xi^3+\xi^2
          \end{array} \right.
\label{eq_a04}
\end{equation}
or, in two-branch form with $\pm$
\begin{equation}
       \beta_{\pm}(\alpha) =  \frac{2}{9}\left(\alpha+\frac{1}{3}\right)
               \left(-1\pm\sqrt{1+3\alpha}\right)-\frac{\alpha}{9}.
\label{eq_a05}
\end{equation}
The cups is at $\alpha=-\frac{1}{3}$ and $\beta=\frac{1}{27}$,
when $\xi=-\frac{1}{3}$.  At the cusp, it is not differentiable:
$d\beta/d\alpha=\infty$ according to Eq. \ref{eq_a04}, but
$\xi=-\frac{1}{3}$ according to Eq. \ref{eq_a05}.

{\bf\em B. Stochastic differential equation with phase transition.}
Stochastic differential equation (SDE)
\begin{equation}
     dx(t) = b(x)dt + \sqrt{2A}\ dB_t
\end{equation}
has a Fokker-Planck equation
\begin{equation}
     \frac{\partial\rho(x,t)}{\partial t}
    = A\frac{\partial^2\rho(x,t)}{\partial x^2}
      -\frac{\partial}{\partial x}\left( b(x)\rho(x,t)
              \right).
\end{equation}
Its stationary solution is
\begin{eqnarray}
      \rho^{ss}(x) &=& C\exp\left(\frac{1}{A}\int b(x)dx \right)
\nonumber\\
          &=& C\exp\left[-\frac{1}{A}\left(
          \frac{x^4}{4}+\frac{x^3}{3}-\frac{\alpha x^2}{2}
                   +\beta x
                \right)\right],
\end{eqnarray}
where $C$ is a normalization factor
\begin{equation}
     C^{-1} = \int_{-\infty}^{\infty}
        \exp\left[-\frac{1}{A}\left(
          \frac{x^4}{4}+\frac{x^3}{3}-\frac{\alpha x^2}{2}
                   +\beta x
                \right)\right] dx.
\end{equation}

Let us denote
\begin{equation}
   \varphi(x) = -\int b(x)dx
              = \frac{x^4}{4}+\frac{x^3}{3}-\frac{\alpha x^2}{2}
                   +\beta x.
\end{equation}
The $\varphi(x)$ has two minima separated by a
maximum.  And the condition for the two minima
being equal is a line in $\alpha\beta$ plane:
\begin{equation}
          \frac{\alpha}{3}+\beta+\frac{2}{27}=0,
\label{maxwell_cut}
\end{equation}
along which we have
\begin{equation}
   \varphi(x) = \frac{1}{4}\left(x+\frac{1}{3}\right)^4
                 -\frac{1}{2}\left(\alpha+\frac{1}{3}\right)
            \left(x+\frac{1}{3}\right)^2 + C',
\label{eq_a012}
\end{equation}
where $C'$ is a constant. Note that the $\varphi(x)$ in
Eq. \ref{eq_a012} is an even function of $\wtx=x+\frac{1}{3}$:
\begin{equation}
     \varphi(\wtx) = \frac{\wtx^4}{4}
             -\left(\alpha+\frac{1}{3}\right)
                  \frac{\wtx^2}{2} + C'.
\label{pfbi}
\end{equation}
This indicates that any cubic system can be
transformed into Landau's canonical energy  form.
Eq. \ref{pfbi} has two minima with equal value when $\alpha>-\frac{1}{3}$; It turns into a single minimum at
$\wtx=0$ when $\alpha<-\frac{1}{3}$. The $\varphi(\wtx)$
is the canonical form of pitch-fork bifurcation \cite{strogatz_book}.  The line in Eq. \ref{maxwell_cut}
is the ``phase separation line'', the location for
Maxwell-like construction.  Its slope
$d\beta/d\alpha = -\frac{1}{3}$ is consistent with
Eq. \ref{eq_a05}.

	In the phase-transition theory language,
$\wtx$ is called an order parameter. Let
$\tau=3\alpha+1$ and $J=27\beta-1$.  Then we have
along the line (\ref{maxwell_cut}) $(\wtx^*)^3-3\tau\wtx^*=0$,
and
\begin{equation}
  \left(\frac{d\ln\wtx^*}{d\ln\tau}\right)_{\tau=0} = \frac{1}{2},
      \ \ \ \textrm{\it i.e.,} \ \wtx^*\propto \tau^{\hat{\beta}}, \
        \hat{\beta} = \frac{1}{2}.
\end{equation}
$\varphi$ in (\ref{pfbi}) is called free energy.
Near $\tau=0$, substituting $\wtx^*\propto \tau^{\hat{\beta}}$, we
have $\varphi\propto\tau^2$.  Then ``heat capacity''
$C=\alpha(\partial^2\varphi/\partial\alpha^2)\propto\tau^{\hat{\alpha}}$
with $\hat{\alpha}=0$.
Furthermore, from $\wtx(\tau,J)$ given by (\ref{beq0}),
let $\chi=\partial \wtx/\partial J
=\big(9\tau-81\wtx^2\big)^{-1}$.  Then
$\chi\propto \tau^{-\gamma}$ with $\gamma=1$.
Finally, for $\tau=0$, Eq. \ref{beq0} becomes
$\wtx^3+J/27=0$. Thus $J\propto \wtx^\delta$
with $\delta=3$.  We note that $\hat{\alpha},\gamma,\delta$ satisfy $2-\hat{\alpha}=\gamma\frac{\delta+1}{\delta-1}$; and $\hat{\beta},\gamma,\delta$ satisfy $\hat{\beta}=\frac{\gamma}{\delta-1}$.

\end{document}